\documentclass[showpacs,preprintnumbers,preprint,amsmath,amssymb,floatfix]{revtex4}
\usepackage[usenames]{color}
\usepackage{epsfig}
\usepackage{amssymb}

\newcommand{\be}{\begin{equation}}
\newcommand{\ee}{\end{equation}}
\newcommand{\bee}{\begin{eqnarray}}
\newcommand{\eee}{\end{eqnarray}}

\newcommand{\piNetaN}{\mbox{$\pi N \! \rightarrow \! \eta N \:$}}

\definecolor{grey}{rgb}{0.9,0.9,0.9}
\definecolor{black}{rgb}{0,0,0}

\begin{document}

\title{ A Missing Link Between Quark-Model Resonant States \\ and Scattering-Matrix Singularities }

\author{S. Ceci, A. \v{S}varc, and B. Zauner}
\affiliation{Rudjer Bo\v{s}kovi\'{c} Institute, \\
Bijeni\v{c}ka c. 54, \\ 
10 002 Zagreb, Croatia\\ 
E-mail: alfred.svarc@irb.hr}
\date{\today}
\vspace*{5.cm}
\begin{abstract}
For last two decades different quark models have predicted diverse, sometimes contradictory collections of 
resonant states. To choose the best among them, the obtained sets had to be compared to available experimental 
values. In the absence of a more thorough understanding, quark-model resonant states have been directly 
identified with scattering-matrix singularities. We demonstrate that these are two closely related, but different 
physical quantities, and offer a model based on the coupled-channel formalism to connect them in an unambiguous 
way. 
\end{abstract} 
\maketitle

\section{Introduction}
The missing-resonance problem, a failure to experimentally confirm a number of unambiguously predicted 
quark-model states (standard, glue enriched, or created from color neutral di- or multi-quark molecules), is for 
quite some time producing a major queasy in the field of hadron spectroscopy \cite{Cap00}. The dilemma has arisen 
whether to suspect the soundness of quark model scheme in general, or the very procedure of connecting 
theoretical predictions with measurable quantities (experimental observables). We track the origin of the problem 
to the latter. Until today, as no better possibilities have been offered, quark-model resonant states have been 
directly compared to scattering-matrix pole parameters, disregarding the fact that the equality between the two 
is valid only if the interaction is neglected; i.e. in the first order perturbation theory. As it has been 
clearly said in ref. \cite{Cap93}, ``a calculation of a spectrum should automatically include some description of 
couplings, as these do affect masses (and vice versa)". However, instead of doing so, the approach in that 
reference has been adopted to treat the problem only as an ``approximate step-by-step process", and calculate the 
first order approximation only.  This standing has been followed practically unanimously. Quark-model resonant 
states are usually compared to Breit-Wigner parameters which serve to quantify experimentally identified 
scattering matrix resonant-like behavior, entirely forgetting about the dressing effects due to the transition 
interaction which are necessarily shifting quark-model resonant state values. Furthermore, yet another unresolved 
issue is for quite some time troubling the physics community: there is a prevailing belief that the Roper 
resonance differs significantly from other resonances (most recently recapitulated in ref. \cite{Roper}), but a 
fully convincing understanding of the phenomenon has never been given. Possible explanations have been ranging 
from Roper resonance being a hybrid state with excited glue \cite{Bar83}, to understanding it as a five quark 
$qqqq\bar{q}$ state which produces a scattering matrix resonant behavior without a standard three-quark pole term 
\cite{Kre00} (dynamic resonance). 

In this paper we propose a way out for both problems.  

Coupled-channel T-matrix formalism (CC$\_ \,$T) \cite{Cut79,Bat98,Vra00} by construction distinguishes between 
scattering-matrix poles and bare Green function (bare propagator) poles. The bare Green function poles, which are 
the subset of CC$\_ \,$T  model fit parameters, can not be detected experimentally. To became observable they 
have to interact. Through the formalism described by resolvent Bethe-Salpeter equation the self-energy term is 
generated; the self-energy term shifts the initial real-value bare propagator poles into the complex energy 
plane; and eventually the measurable complex scattering-matrix poles are generated as dressed Green function 
poles. 

The CC$\_ \,$T formalism we describe, has been known for decades, but until present moment no physical 
interpretation has been given to bare propagator pole parameters. For the first time, we offer one possibility. 
\newpage \noindent
We propose: \\
the position of a bare Green function pole is to be identified with the mass of a quark-model resonant state 
(QMRS); the imaginary part of the scattering-matrix pole (SMP), which is created when the interaction effects 
shift QMRS into the complex energy plane, is to be correlated with its decay width. 
%
%

 Such an identification simultaneously solves both problems: establishes a missing link between QMRS and SMP 
offering a better control over the missing-resonance problem, and at the same time creates a mechanism how to 
distinguish between genuine scattering-matrix resonant states (SMRS) and dynamically generated ones. By accepting 
this assumptions, we are able to: a) identify which QMRS are needed to explain a chosen collection of 
experimental data;  b) determine a nature of a given SMRS (genuine or dynamic).

We believe that the interpretation we propose is a correct one because coupled-channel formalism ``simulates 
reality''. Namely, in the  ``real world'', bound and resonant states of two or more quarks and antiquarks are 
considered to be a basis of meson-nucleon interactions. These states, usually described by a quark-model 
Hamiltonian with its proper values being their bare masses, are not physical observables. To become measurable 
quantities, these states first have to interact, and only when the interaction is introduced, we can talk about 
their decay width in its standard physical meaning. The coupled-channel model we use does the same. It starts 
with bare propagator poles, and through the interaction shifts them into measurable SMPs. So again we end up with  
two well defined, correlated, but distinguished physical quantities, exactly as we claim the situation in the 
real world is.  
\section{Formalism}

\begin{center}
\begin{figure}[!h]
\includegraphics[width=6.5cm]{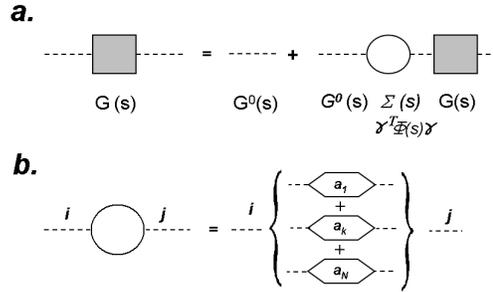}\\
 \caption{Resolvent Bethe-Salpeter equation and self-energy term.}
\label{figure1}
\end{figure}
\end{center}

For the convenience of the reader, we shall recapitulate the essence of coupled-channel T-matrix formalism. We 
use the Carnegie Melon-Berkeley (CMB) approach \cite{Cut79}. CMB model is a separable coupled-channel 
partial-wave analysis with two main ingredients: bare resonant propagator ${G}_{0} (s)$ - the diagonal matrix in 
\emph{resonant} indices incorporating real first-order poles and the channel-propagator ${\Phi} (s)$ -  the 
diagonal matrix in \emph{channel} indices with matrix elements $\phi (s)$, which takes care of other non-pole 
singularities. The solution of the problem, the dressed resonant propagator ${G}(s)$, is given by the resolvent 
(Bethe-Salpeter) equation \mbox{${G}^{-1} (s) = {G}^{-1}_{0} (s) - {\Sigma} (s)$}, See Fig. \ref{figure1}.a. The 
self-energy term ${\Sigma} (s)$, as shown in Fig. \ref{figure1}.b,  is built from the channel propagator as 
${\gamma^{T}}  \Phi  \gamma$, therefore the model manifestly satisfies unitarity.
Both, non-square parameter matrix $\gamma$ and values of the real bare propagator poles $s_0$,  are concurrently  
obtained from the least-square fit of unitary-normalized partial-wave   ${ T = \sqrt{\mathrm{Im}\Phi} \, \gamma 
\,G \, \gamma^T \sqrt{\mathrm{Im}\Phi}}$ to experimental partial-wave data. 
\newpage
The formal way to present the main hypothesis of this article is to discuss the graphical representation of 
Bethe-Salpeter equation given in Fig. \ref{figure1}. 
From Fig. \ref{figure1}.a it is clear that measurable SMPs will be given by the poles of the dressed Green 
function ${G}(s)$. 
The resolvent equation states that the dressed propagator poles are generated from the poles of the bare Green 
function $ {G}_{0} (s)$ through the process of ``dressing" them with the self-energy term $ {\Sigma} (s)$. And 
this is the term which, as depicted in Fig. \ref{figure1}.b, quantifies the transition interaction when bare 
propagator poles undergo through all possible intermediate states $a_k$. Consequently, we end up with two 
connected sets of poles: bare and dressed Green function poles; different but directly related through the 
interaction - just like  in the ``real world". Therefore it seems natural to connect real-number bare propagator 
pole $s_0$ with QMRS; and the imaginary part of the SMP, produced when complex self-energy $ {\Sigma} (s)$ pushes 
QMRS into the complex energy plane, with the decay width. Let us observe that, in addition to creating the 
imaginary part, bare propagator poles (QMRS masses) are shifted as well. 

 By accepting our assumption quite a number of unresolved issues is put into order: SMPs are set apart, but 
directly linked to the QMRS; the nearest QMRS responsible for a specific SMP can be identified; the mechanism is 
opened by which the interaction can push the ``reasonable" QMRS into the complex energy domain \emph{not 
accessible} to present experiments (QMRS gets lost); the mechanism is opened by which the SMPs can be created not 
from QMRS, but otherwise.  
\section{Results}

We illustrate the manner how our hypothesis works on a very simple three channel model.

 Even before showing the results, we want to warn the reader that the simplicity of the model, i.e. the fact that 
we are effectively using only three out of at least seven accessible and contributing channels, will produce only 
qualitative results. Complexity of the coupled-channel model (simultaneous mixing of all channels) requires 
considerable number of parameters, and we expect that the absence of constraining data in more than two channels 
will necessarily produce instabilities in obtained fitting solutions. We expect to achieve a better quantitative 
agreement between bare propagator pole position and quark-model results  once the stability of the fit solutions 
is ensured by including other available inelastic channels. What we do need in future, and what we are constantly 
stressing on numerous occasions (last time in \cite{Sva04}), is to have the constraining data in as many channels 
as possible. The quality of the data is of importance, but the abundance of constraining channels is what counts. 
\begin{figure}[!h]
\includegraphics[width=5cm]{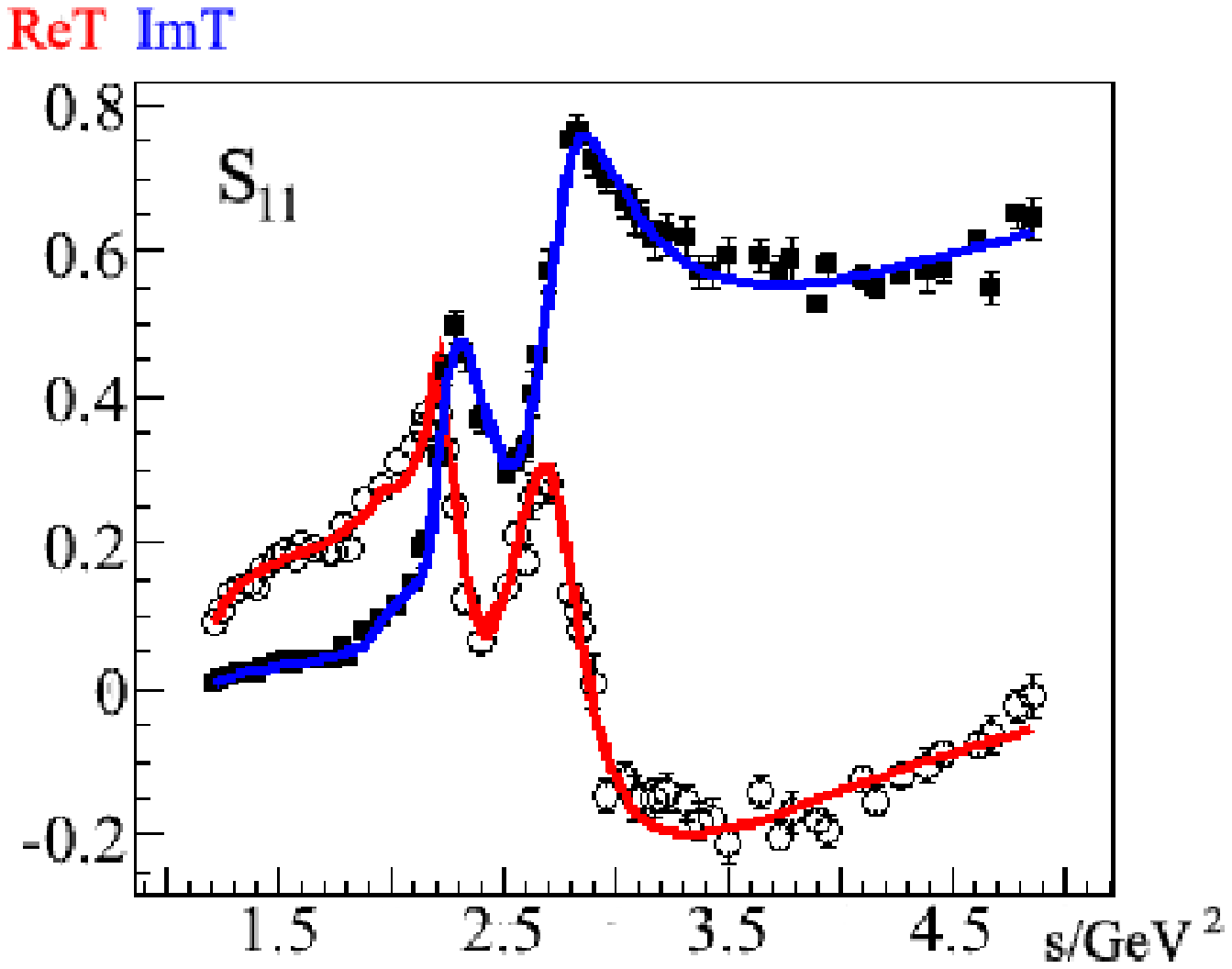} \hspace*{2cm}
\includegraphics[width=5cm]{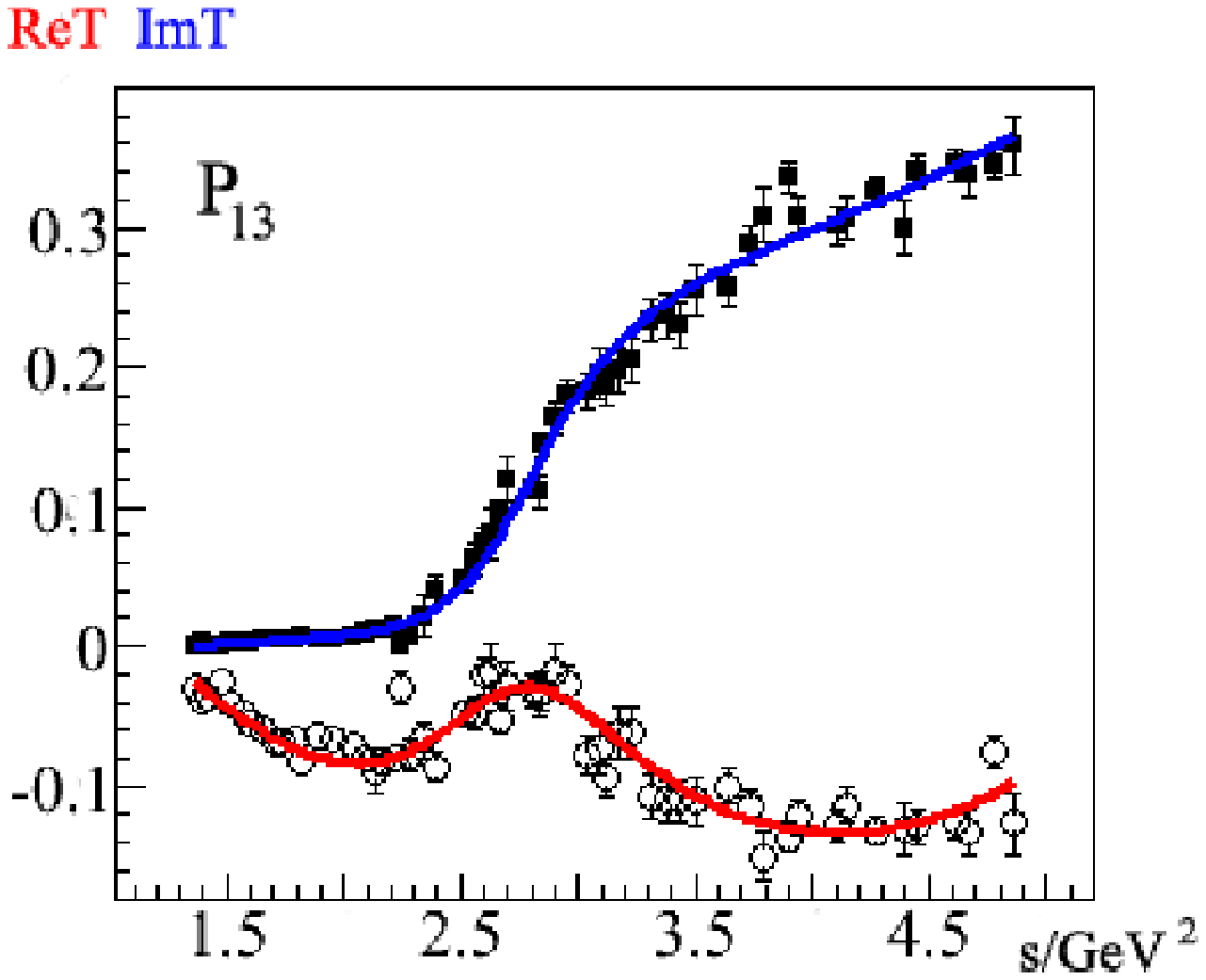} \vspace*{0.2cm} \\
\includegraphics[width=5cm]{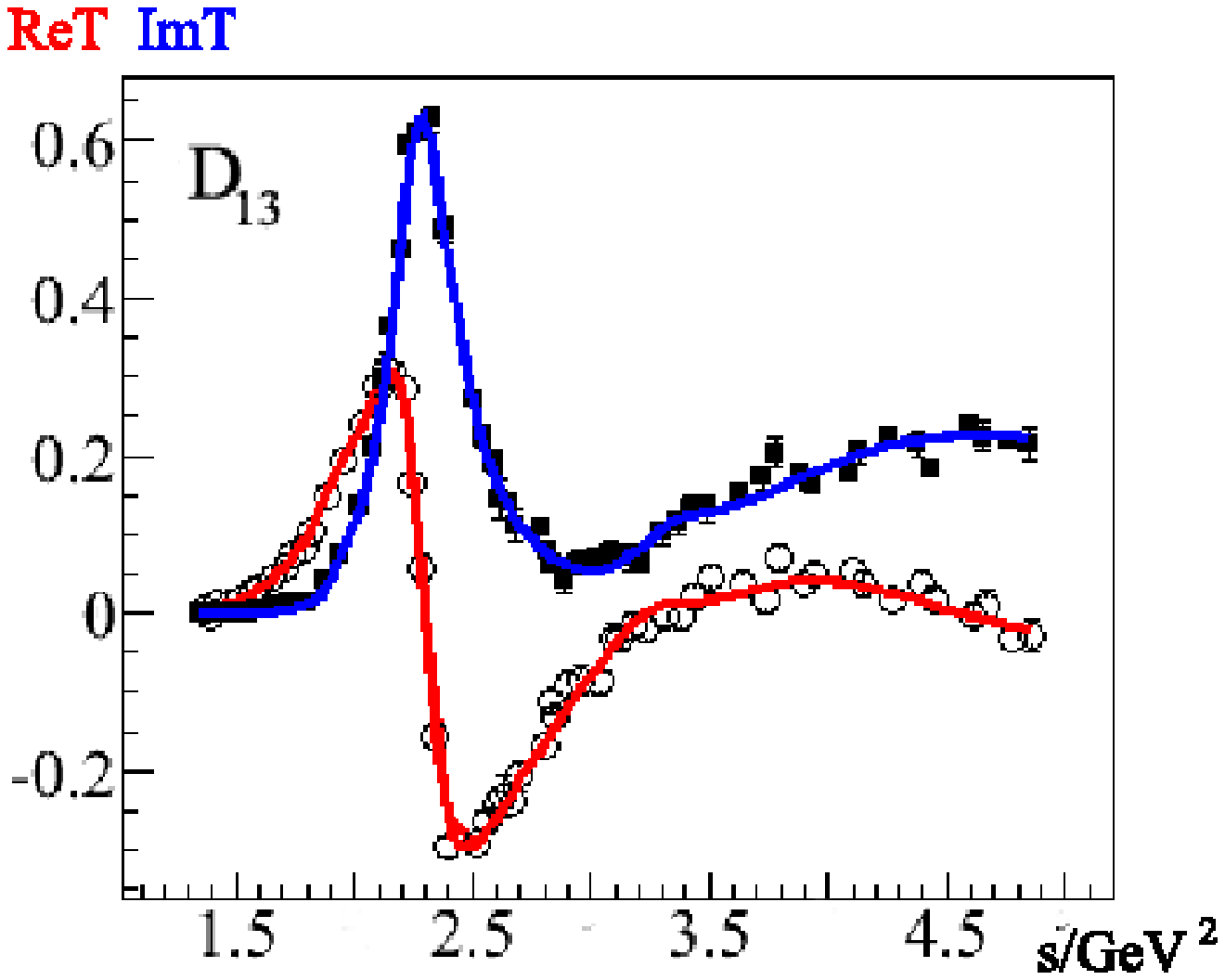} \hspace*{2cm}
\includegraphics[width=5cm]{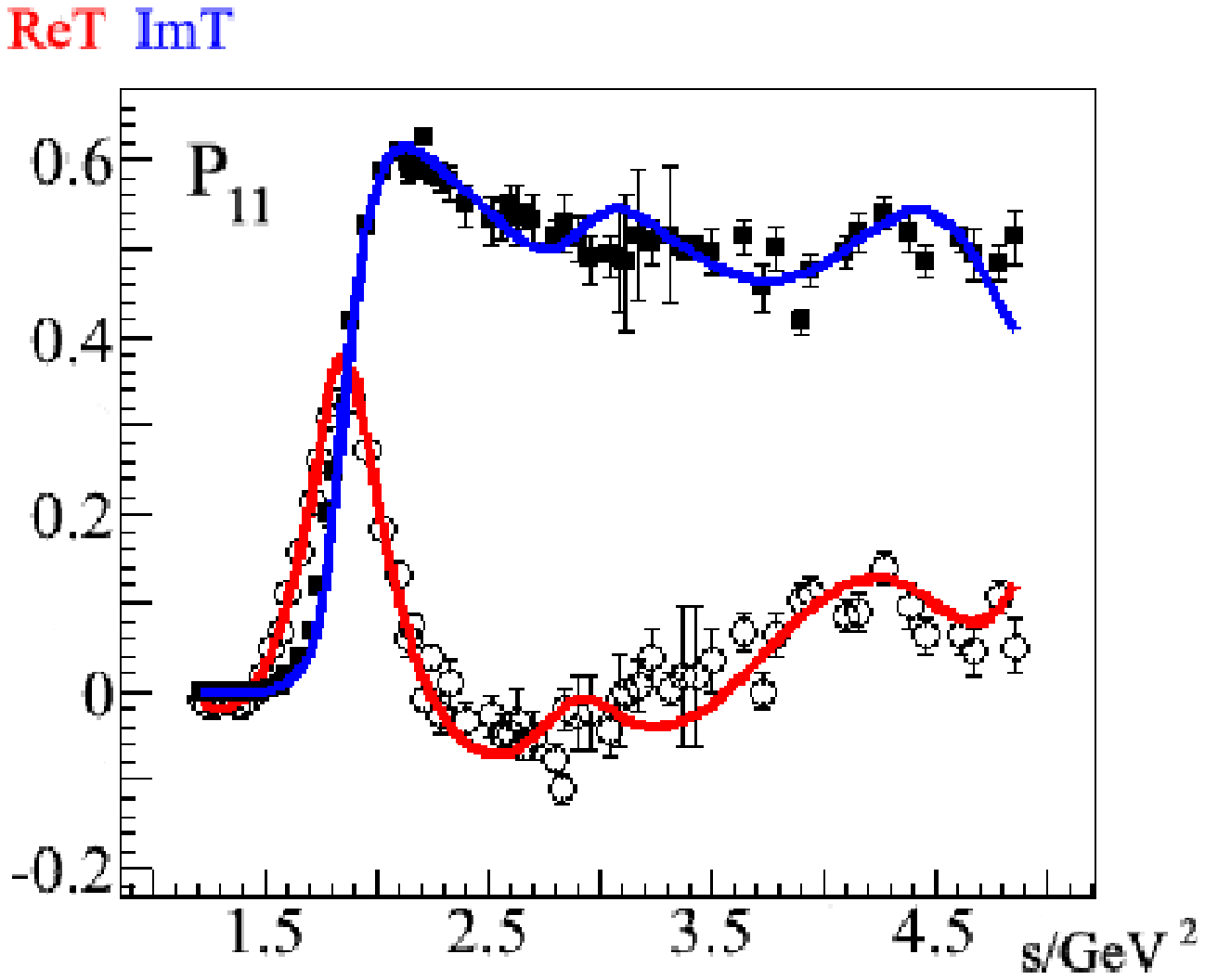} 
 \caption{The obtained curves for $\pi N$ elastic channel. Empty squares and red lines denote the real part of 
the scattering matrix and full squares and blue lines denote the imaginary one. }
\label{figure2}
\end{figure}
\begin{figure}[!h]
\includegraphics[width=5cm]{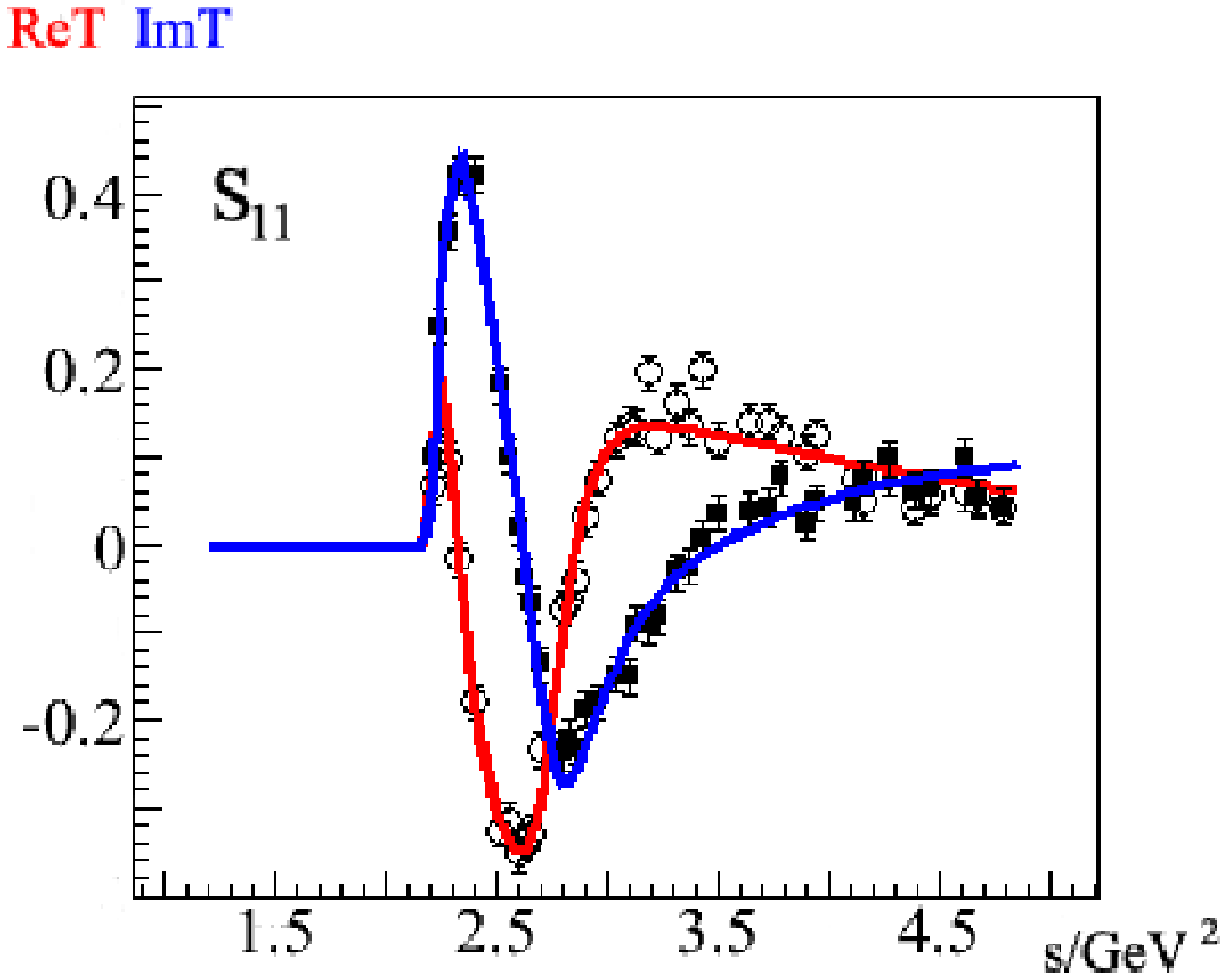} \hspace*{2cm}
\includegraphics[width=5cm]{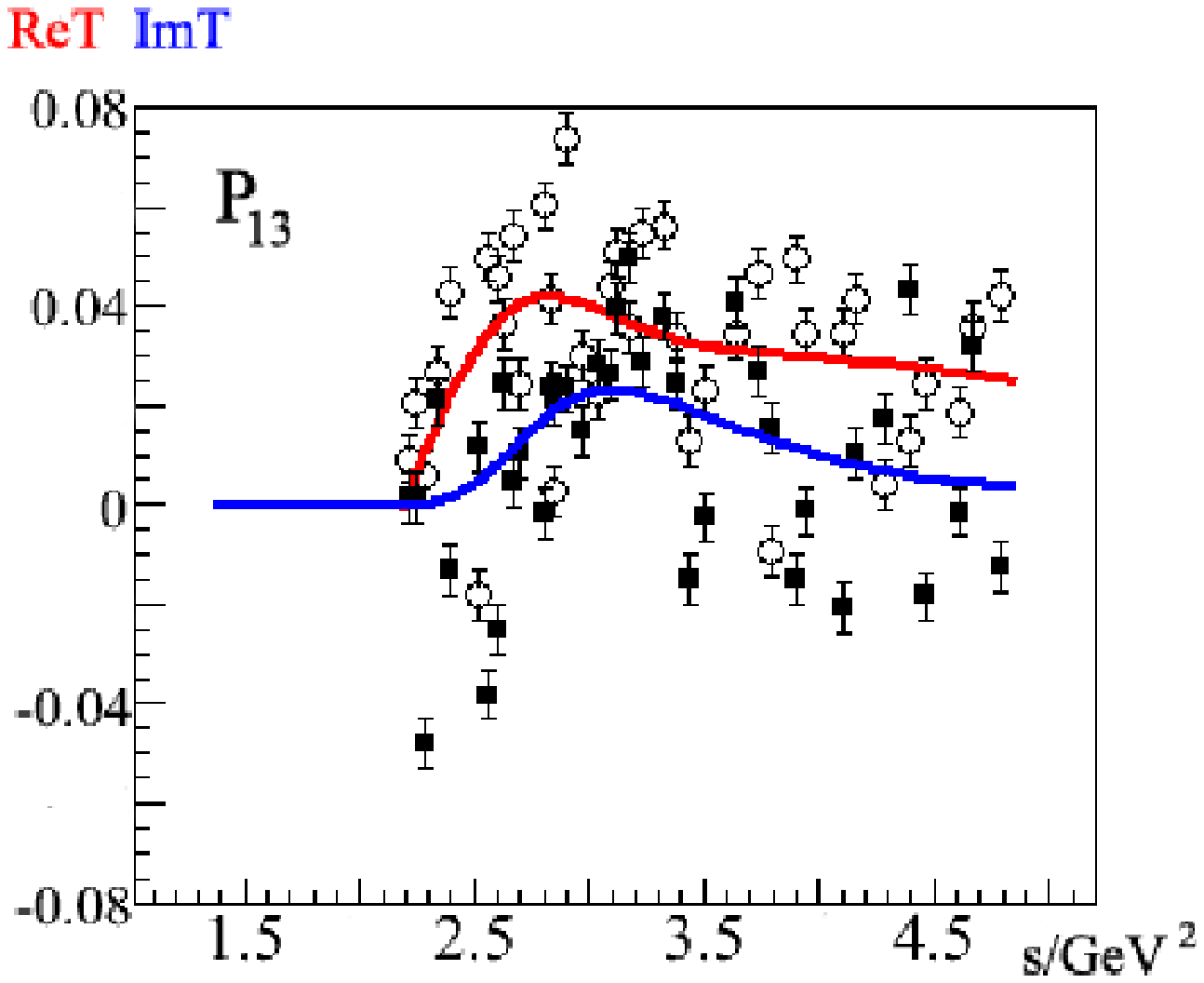}  \vspace*{0.2cm} \\
\includegraphics[width=5cm]{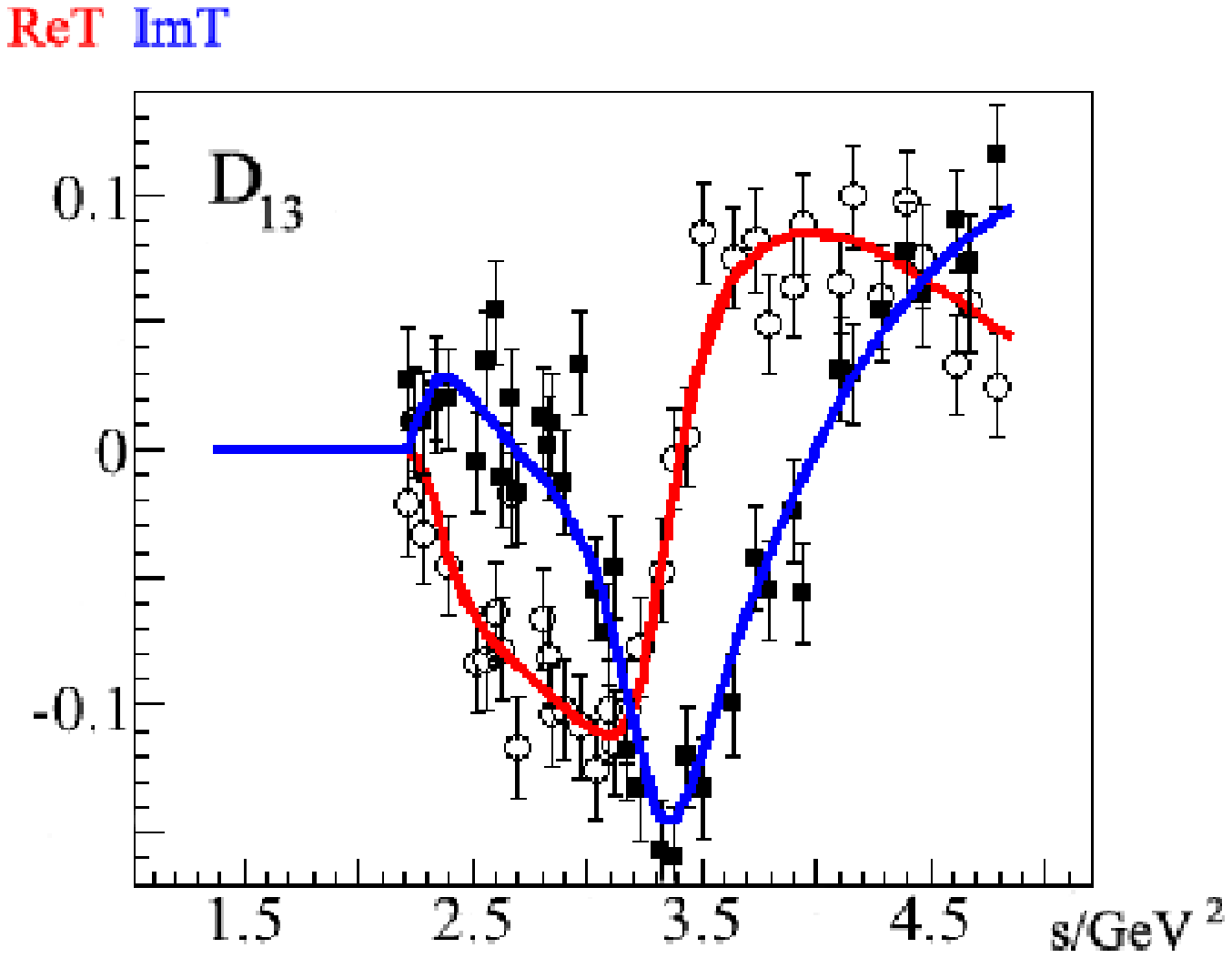} \hspace*{2cm}
\includegraphics[width=5cm]{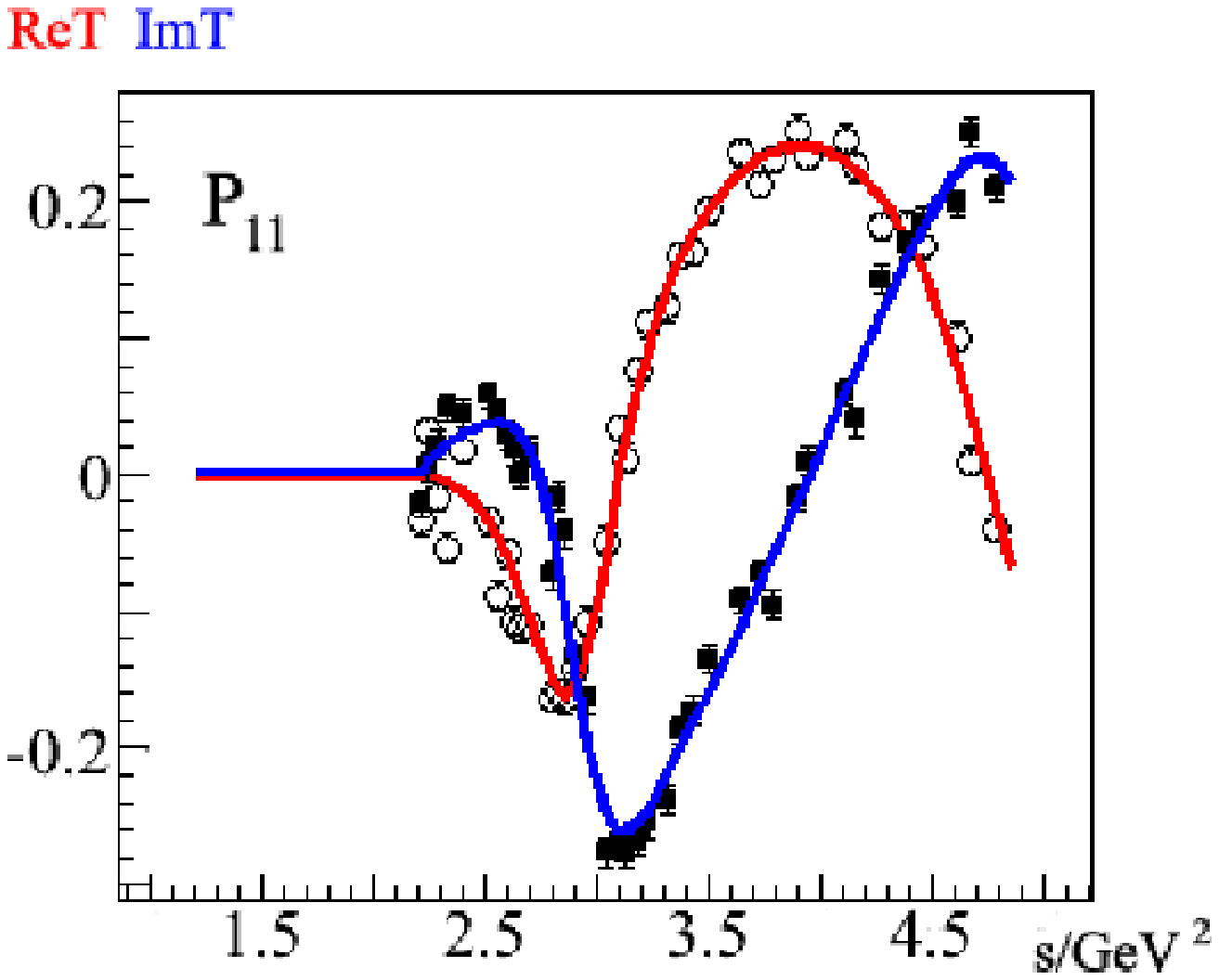} 
 \caption{The obtained curves for \piNetaN channel. }
\label{figure3}
\end{figure}

The first four partial waves in I=1/2 channel (S$_{11}$,  P$_{11}$,  P$_{13}$ and  D$_{13}$) were analyzed. We 
use a model with three channels: two physical two-body channels $\pi N$ and $\eta N$, while the third, effective 
channel represents all remaining two- and three-body processes in a form of a two-body process. \\
 For the $\pi N$ elastic partial waves we used the VPI/GWU single-energy solutions \cite{GWUWEB,Arn04}.
 \\
For the \piNetaN partial-wave data we used the coupled-channel amplitudes from Batini\'{c} \emph{et al.}  
\cite{Bat98}, but instead of using smooth theoretical curves, we constructed the data points by normally 
distributing the model input (see ref. \cite{Cec06}).
\\ \noindent
Fitting strategy was taken over from ref. \cite{Cec06}.
\\ \noindent
The obtained curves correctly reproduce all input partial wave data and are given in Figs. \ref{figure2} and 
\ref{figure3} for $\pi N$ elastic and \piNetaN process. 
\\ \noindent
The obtained bare propagator and scattering-matrix poles are collected in Table I. 

We show two groups of results: two lowest negative and two lowest positive parity partial waves. The first column 
contains bare propagator poles produced by the model (upper line, normal text) and one out of several available 
sets of quark-model masses (lower line, italics) \cite{Cap86}. Second column contains scattering-matrix poles 
produced by the CC$\_ \,$T model (upper line, normal text) and their standard experimental values (lower line, 
italics) \cite{PDG06}. 
\begin{table*}[!ht]
\caption{Bare propagator and scattering-matrix poles.} 
\begin{tabular}{c | c c c l | c c c c c}
 \hline \hline
 \multicolumn{1}{c |}{} & \multicolumn{4}{c|}{\rm \ \ bare \ propagator \ poles w$=\sqrt{s_0}$ } & 
\multicolumn{5}{c}{\rm \ \ scattering-matrix \ poles - our model \ \ } \\ [-1.ex]
\multicolumn{1}{c |}{} & \multicolumn{4}{c|}{\it \ \ quark \ model \ masses - ref. \cite{Cap86} } & 
\multicolumn{5}{c}{\it \ \ scattering-matrix \ poles - ref. \cite{PDG06} \ \ } \\ [-1ex]
\multicolumn{1}{c |}{\rm \ \  partial \ } & \multicolumn{4}{c |}{\rm (GeV)} & \multicolumn{5}{c}{\rm (GeV)} \\
\cline{2-10}

{\rm wave} & No.1   & No.2   & No.3   & \ \ No.4, $\cdots$   & No.1   & No.2   & No.3   &   No.4  &   No.5  \\
{} & (w) & (w)  & (w) & \ \ (w)   & $\textstyle \binom{\rm Re\, w}{\rm -2\, Im\, w}$ & $\textstyle \binom{\rm 
Re\, w}{\rm -2\, Im\, w}$ & $\textstyle \binom{\rm Re\, w}{\rm -2\, Im\, w}$  & $\textstyle \binom{\rm Re\, 
w}{\rm -2\, Im\, w}$ & $\textstyle \binom{\rm Re\, w}{\rm -2\, Im\, w}$ \\ [0.5ex]
 \hline \hline 
S$_{11}^{-}$ & 1.518 & 1.651 & 1.912 & - & $\textstyle\binom{1.51}{0.1}$ & $\textstyle\binom{1.65}{0.14}$ & 
$\textstyle\binom{1.91}{0.95}$ & -  & -  \\ 
{} & {\it \footnotesize 1.510} & {\it \footnotesize 1.585} & {\it \footnotesize 1.945} & {\it \scriptsize 2.030, 
$\cdots$ } & $\textstyle\binom{\it 1.51}{\it 0.10}$ & $\textstyle\binom{\it 1.65}{\it 0.14}$ & 
$\textstyle\binom{\it 2.15}{\it 0.35}$ & - & -\\
D$_{13}^{-}$ & 1.477 & 1.663 & 1.934 & - & \ $\textstyle\binom{1.40}{0.15}$ & $\textstyle\binom{1.51}{0.11}$ & 
$\textstyle\binom{1.82}{0.16}$ & - & -\\
{} & {\it \footnotesize 1.495} & {\it \footnotesize 1.625} & {\it \footnotesize 1.960} & {\it \scriptsize 2.055, 
$\cdots$ }& \ $\textstyle\binom{\it 1.51}{0.11}$ & $\textstyle\binom{\it 1.68}{\it 0.10}$ & $\textstyle\binom{ 
\it 1.88-2.05}{\it 0.16-0.20}$ & - & -\\
\hline
P$_{13}^{+}$ & 1.909 & 2.484 & - & - & $\textstyle\binom{1.65}{0.36}$ & $\textstyle\binom{2.47}{0.18}$ & - & - & 
- \\
{} & {\it \footnotesize 1.795} & {\it \footnotesize 1.870} & {\it \footnotesize 1.910} & {\it \scriptsize 1.950, 
2.030, $\cdots$ } & $\textstyle\binom{\it 1.675}{\it 0.115-0.275}$ & $\textstyle\binom{\it 1.90^*}{\it 0.498^*}$ 
& - & - & - \\
P$_{11}^{+}$ & \ 0. 960 \ & \ 1.854 \ & \ 2.018 & \  2.759 \ & $\textstyle\binom{1.10}{0.00}$ & 
\colorbox{grey}{$\textstyle\binom{1.35}{0.16}$} & $\textstyle\binom{1.70}{0.10}$  & \ 
$\textstyle\binom{2.00}{0.60}$ & \ \ \ $\textstyle\binom{2.20}{0.20}$ \\
{} & \ {\it \footnotesize 0. 960} \ & \ {\it \footnotesize 1.540} \ & \ {\it \footnotesize 1.770} \  & \ {\it 
\scriptsize 1.880, 1.975, 2.065, $\cdots$ }  & - & \colorbox{grey}{$\textstyle\binom{\it 1.365}{\it 0.19}$} & 
$\textstyle\binom{\it 1.72}{\it 0.23}$  & \ $\textstyle\binom{\it 2.12}{\it 0.24}$ & - \\
\hline\hline
\end{tabular}
\label{table1} 
\end{table*}  

\begin{figure}[!h]
\includegraphics[width=6cm]{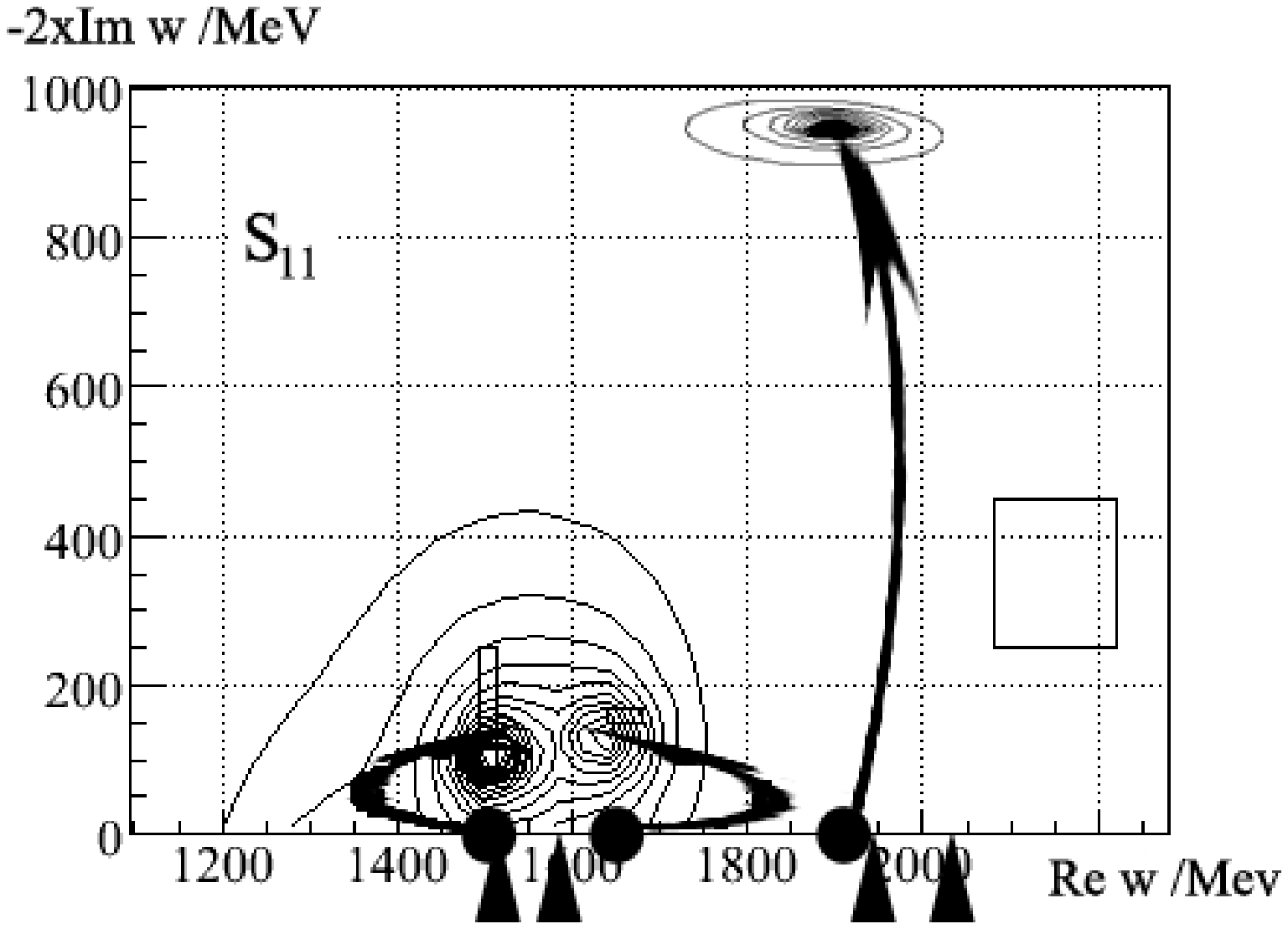} \hspace*{1.cm}
\includegraphics[width=6cm]{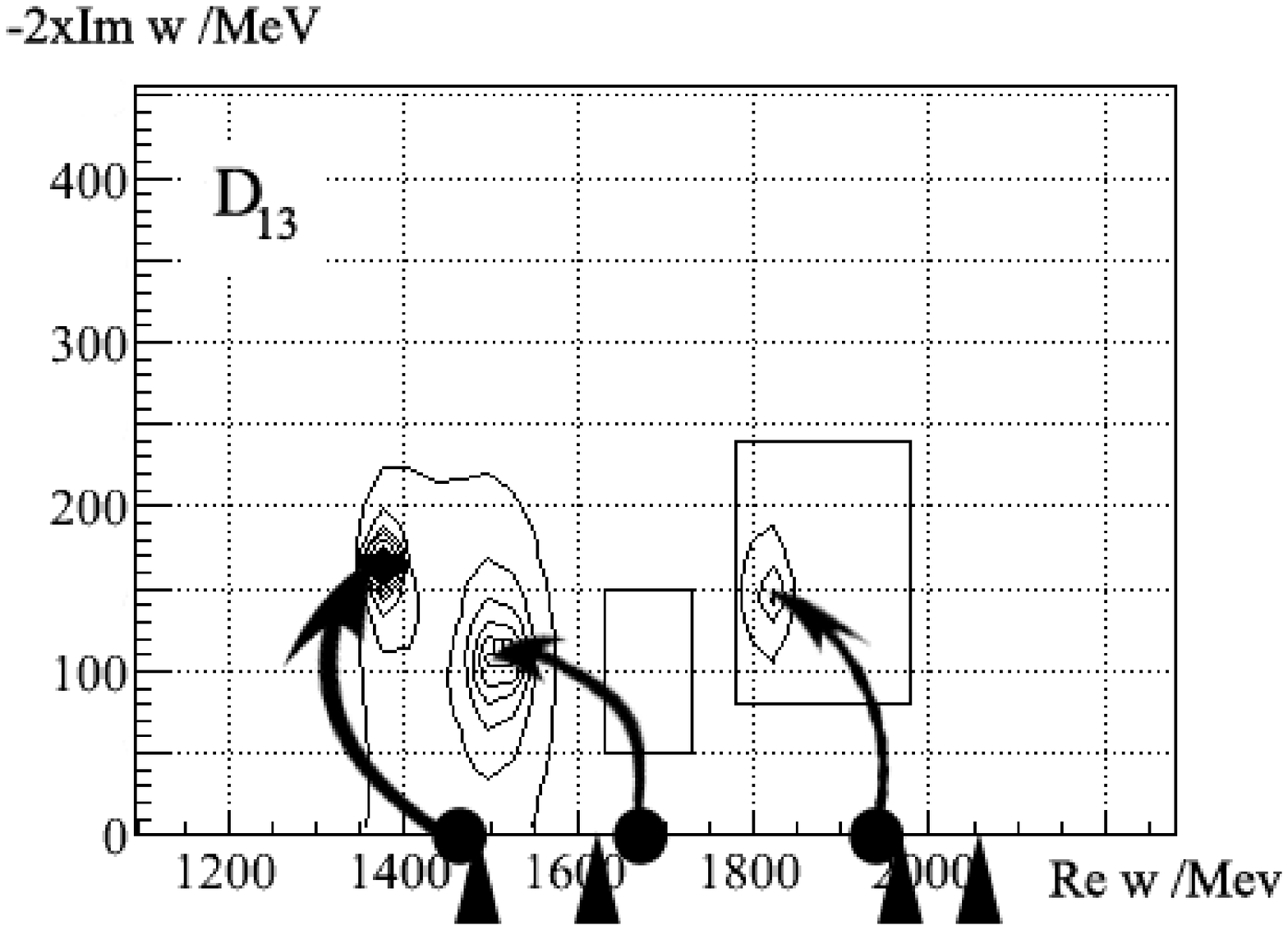} 
 \caption{scattering-matrix singularities and bare propagator pole positions for the two lowest negative parity 
states. Full dots denote bare propagator pole positions, triangle arrows denote the few lowest quark-model 
resonant state masses of refs. \cite{Cap86}.  }
\label{figure4}
\end{figure}

First group of results, the two lowest {negative} parity partial waves S$_{11}$ and D$_{13}$, pretty well 
confirms our claims. As can be seen in Table I and in Fig. \ref{figure4}, all three bare propagator poles for 
both partial waves can be naturally identified with lowest QMRS of refs. \cite{Cap86}.  We do see some 
discrepancies in mass position, but each required bare propagator pole does qualitatively correspond to a 
particular QMRS, and all lowest QMRS have found their bare propagator counter partners.  

The obtained  CC$\_ \,$T scattering-matrix pole positions correspond reasonably well to the experimental values 
reported in ref. \cite{PDG06}. The only disagreement, the unexpected position of the third SMP of the S$_{11}$ 
partial wave (too far in the complex energy plane), is again a consequence of the fit-results instability, and is 
expected to disappear with including more channels. All three experimentally detected  SMPs for the D$_{13}$ 
partial wave are reproduced, but the lowest two are somewhat shifted in mass. 

\begin{figure}[!h]
\includegraphics[width=6cm]{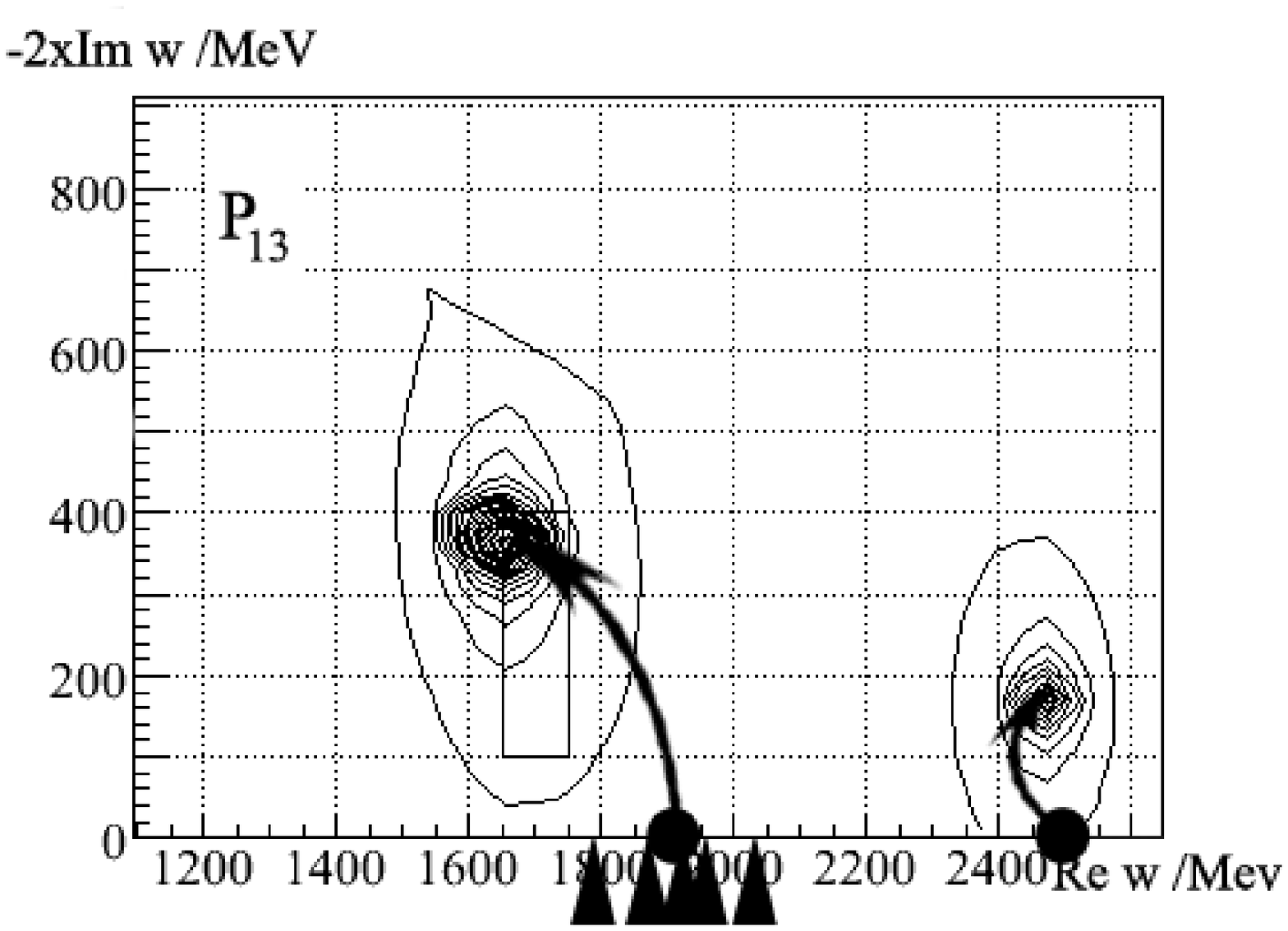} \hspace*{1.cm}
\includegraphics[width=6cm]{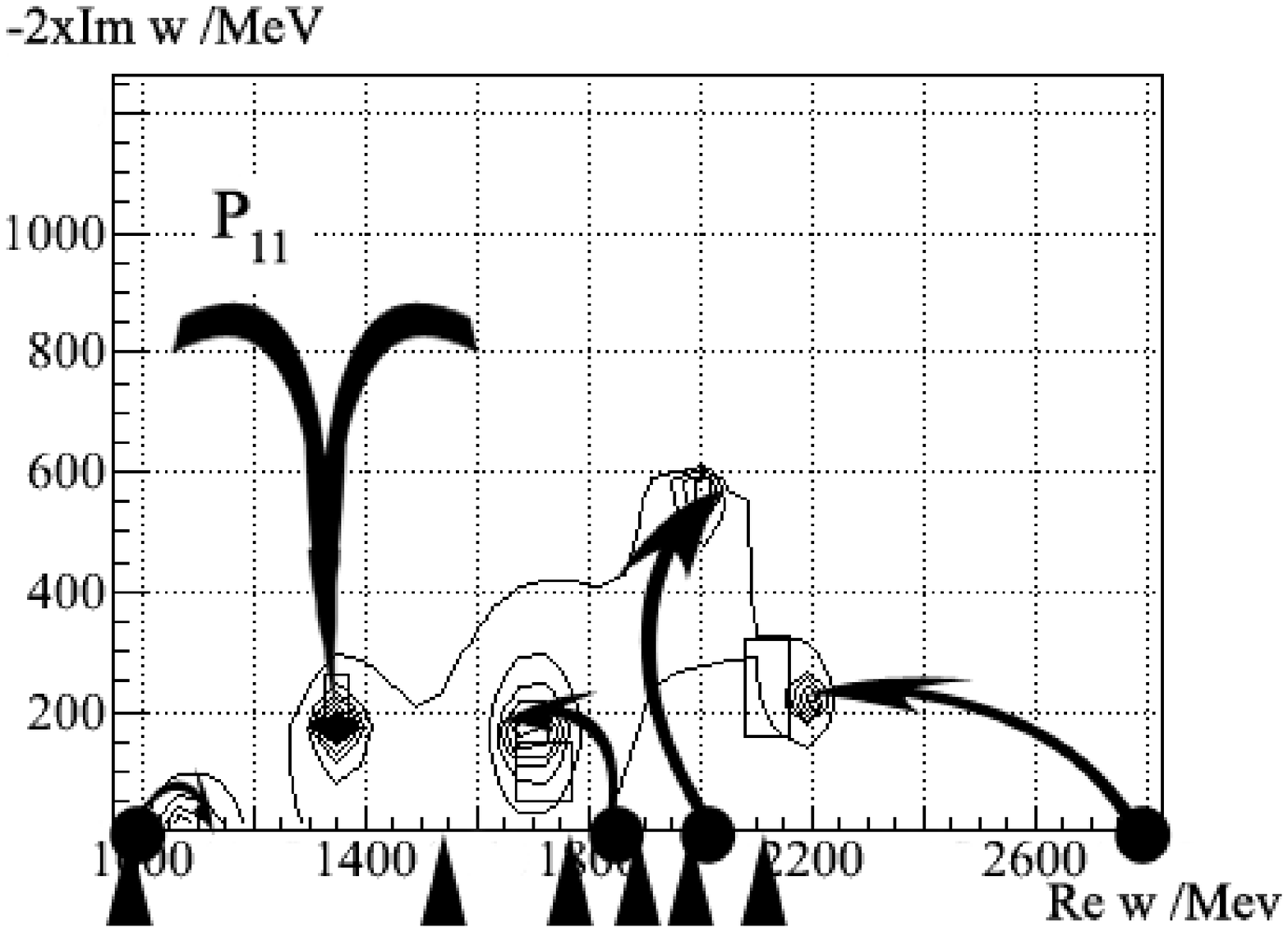} 
 \caption{scattering-matrix singularities and bare propagator pole positions for the two lowest positive parity 
states.}
\label{figure5}
\end{figure}

The influence of the interaction upon bare propagator poles, and their ``journey" from the initial QMRS to the 
final SMP positions, is symbolically visualized in Figs. \ref{figure4} and \ref{figure5}. In the world without 
interaction $\gamma$ matrices vanish, we have no ``dressing", and bare propagator and scattering-matrix poles are 
identical. In the real world, in the world with interaction, the $\gamma$ matrices are non-vanishing, and are 
obtained by fitting the partial wave data. Arrows represent the way how bare propagator poles travel from the 
world without interaction ($\gamma$=0) to the real world scattering-matrix singularities  ($\gamma \neq 0$).

Next group of results, the two lowest {positive} parity partial waves P$_{13}$  and P$_{11}$, is still consistent 
with the hypothesis of the article, but some problems appear. As can be seen in Table I and in Fig. \ref{figure5} 
only one out of five QMRS of ref. \cite{Cap86} for the P$_{13}$ partial wave is identified with the bare 
propagator pole, while other states remain yet to be identified. The second required bare propagator pole should 
either be identified with one of the higher lying QMRS, or will be shifted downwards when results of the fit 
stabilize. 

The notoriously problematic P$_{11}$ partial wave, however, remains a troublemaker as in majority of theoretical 
considerations.  First, we needed four bare propagator poles in order to achieve acceptable fit to the input 
data.   Having in mind that we should identify bare propagator poles with \emph{all} quark-model states (resonant 
\emph{and} bound), we have fixed the value of the first bare propagator pole to the mass of the sub-threshold 
nucleon pole, and left the remaining three poles unconstrained. As shown in Table I, all experimentally 
determined SMPs \cite{PDG06} are reproduced. So far so good.

 Problems start with the identification of QMRS with bare propagator pole position. In ref. \cite{Cec06} we have 
demonstrated that the presence of inelastic channels directly produces the { \it N(1710)} P$_{11}$ SMP, and in 
Fig. \ref{figure5} we show that it is generated by dressing the 1.854 GeV bare propagator pole. This pole can be 
directly associated with one of the quark-model states of ref. \cite{Cap86}, either 1.770 or 1.880. The nucleon 
state is producing an insignificant, sub threshold and experimentally inaccessible pole at 1.1 GeV; remaining two 
poles at 2.018 and 2.759 produce SMP of 2.2 GeV which can be identified with poorly determined \emph{N(2100)} 
P$_{11}$, and an experimentally not yet established state at 2 GeV.

However, our model with constraining data in only two channels shows two very interesting features for P$_{11}$ 
partial wave: {\it i) no bare propagator pole which would correspond to the 1.540 quark-model state is needed; 
ii) one of experimentally confirmed SMPs, namely the N(1440) P$_{11}$ state - Roper resonance, is not produced by 
any nearby bare propagator pole as it was the case for all other scattering-matrix poles; it is generated 
differently.} 

The CMB model in conjunction with our interpretation of physical meaning of bare propagator poles offers us a 
natural way to characterize the nature of scattering-matrix resonant state. We propose a criteria: the 
\emph{genuine} SMRS is a state which is produced by a nearby bare propagator pole; the \emph{dynamic} SMRS is a 
state which is created out of distant bare propagator poles through the interaction mechanism itself. Keeping in 
mind the often expressed belief that the Roper resonance differs significantly from other resonances (recently 
summarized in \cite{Roper}), we agree that it might not be a three quark state at all  \cite{Kre00}.  In our 
model, Roper resonance turns out to be a \emph{dynamic }scattering-matrix resonant state. The advantage of our 
model with respect to the approach of ref. \cite{Kre00} is that we do not have to go through a cumbersome  
procedure of adding an extra pole term to the full K-matrix coupled-channel effective Lagrange model and 
demonstrate its superfluousnes. Our straightforward criteria: presence or absence of a nearby bare propagator 
pole gives us immediately the decisive answer. Similarly as in ref. \cite{Kre00}, we in our model do find a 
scattering-matrix pole in the correct energy regime, but that singularity is not produced by a nearby bare 
propagator pole. In other words, we have no need for such an entity as the ``Roper quark-model resonant state".  
\section{Conclusions}

For the first time quark-model resonant states are distinguished from scattering-matrix poles, and are directly 
identified with a set of fitting parameters (\mbox{QMRS $\Leftrightarrow$ bare propagator poles} of the 
coupled-channel T-matrix model). 

The assumption is tested on a simple three-channel coupled-channel model for the first few partial waves and we 
show: 

The bare propagator poles nicely correspond to the low-lying quark-model resonant states.

The \emph{N(1440)} P$_{11}$ resonance (Roper resonance) is a scattering-matrix pole \emph{not} produced by a 
nearby bare propagator pole. Therefore, for the time being, we considered it to be a dynamic resonant state.

The \emph{N(1710)} P$_{11}$ resonance is a genuine scattering-matrix resonant state produced by a nearby bare 
propagator pole which can be identified with a quark-model resonant state. However, its final positioning awaits 
new data in other inelastic channels.   

The lowest P$_{11}$ quark-model resonant state of ref. \cite{Cap86} could not be identified when only $\pi N$ 
elastic and \piNetaN  experimental data are used, so we wonder which data and from which process will confirm its 
existence.

\end{document}